\documentclass[letterpaper, 10 pt, conference]{ieeeconf}  % Comment this line out if you need a4paper
\usepackage[spanish]{babel}
\usepackage[utf8]{inputenc}
\usepackage{listings}% http://ctan.org/pkg/listings
\usepackage{caption}
\usepackage{DejaVuSans}
\usepackage{DejaVuSansMono}
\usepackage{amssymb}
\usepackage{amsmath}

\usepackage{subcaption}
\usepackage{graphicx}

\usepackage{alltt}

\usepackage{fancyvrb}

\usepackage[table,xcdraw]{xcolor}
\usepackage{multirow}

\usepackage[T1]{fontenc}
\usepackage{fancyvrb}
\usepackage{listings}
\usepackage[scaled=.55]{beramono}    
\fvset{baselinestretch=0.5}
\DeclareCaptionFormat{listing}{\rule{\dimexpr0.9\columnwidth+17pt\relax}{0.35pt}\par\vskip1pt#1#2#3}
\newsavebox{\FVerbBox}
\usepackage[most]{tcolorbox}
\tcbset{
    frame code={}
    center title,
    left=0pt,
    right=0pt,
    top=0pt,
    bottom=0pt,
    colback=gray!10,
    colframe=white,
    width=0.45\textwidth,
    enlarge left by=0mm,
    boxsep=5pt,
    arc=0pt,outer arc=0pt,
    }

\IEEEoverridecommandlockouts                              % This command is only needed if 
\title{\LARGE \bf
Real-time Linux communications: an evaluation of the Linux communication stack for real-time robotic applications
}
\author{\textbf{Carlos San Vicente Gutiérrez}, 
    \textbf{Lander Usategui San Juan}, \\
    \textbf{Irati Zamalloa Ugarte}, 
    \textbf{Víctor Mayoral Vilches} \\
    Erle Robotics S.L. \\
    Vitoria-Gasteiz, \\
    Álava, Spain \\
}

\usepackage{blindtext}
\usepackage{graphicx}
\usepackage{listings}% http://ctan.org/pkg/listings
\usepackage{xcolor}

\usepackage{hyperref}
\hypersetup{
    colorlinks=true,
    linkcolor=blue,
    filecolor=magenta,      
    urlcolor=cyan,
    citecolor=blue,
}

\usepackage[normalem]{ulem}

\begin{document}
\maketitle
%\tableofcontents
%\thispagestyle{empty}
%\pagestyle{empty}

%%%%%%%%%%%%%%%%%%%%%%%%%%%%%%%%%%%%%%%%%%%%%%%%%%%%%%%%%%%%%%%%%%%%%%%%%%%%%%%%
\begin{abstract}
%Linux is becoming the preferred operative solution for robotics in embedded systems. 
%Thanks to the work of the Real-time Linux (RTL) project, Linux is becoming a real-time operative system preserving all the advantages of a general purpose operative system (GPOS). 

As robotics systems become more distributed, the communications between different robot modules play a key role for the reliability of the overall robot control. In this paper, we present a study of the Linux communication stack meant for real-time robotic applications. We evaluate the real-time performance of UDP based communications in Linux  on multi-core embedded devices as test platforms. We prove that, under an appropriate configuration, the Linux kernel greatly enhances the determinism of communications using the UDP protocol. Furthermore, we demonstrate that concurrent traffic disrupts the bounded latencies and propose a solution by separating the real-time application and the corresponding interrupt in a CPU.

%Using the Linux kernel stack for real-time Ethernet communications would greatly extend the use of communications middleware such as Data Distribution Service (DDS) used in safety critical robotics applications.\\

%In this work we present an evaluation of the real-time performance of the Linux Network stack comparing a vanilla Linux kernel and a real-time kernel based on PREEMPT-RT. We test different configurations under different load conditions. Finally, based on our results we conclude that it is possible to achieve communications with a proper configuration and under certain conditions.   

% robotics systems tend to be more distributed, reliable communications with sensors and actuators are a fundamental part of the system. 
\end{abstract}

%%%%%%%%%%%%%%%%%%%%%%%%%%%%%%%%%%%%%%%%%%%%%%%%%%%%%%%%%%%%%%%%%%%%%%%%%%%%%%%%
\section{Introduction}
\label{introduction}

The Ethernet communication standard is widely used in robotics systems due to its popularity and reduced cost. When it comes to real-time communications, while historically Ethernet represented a popular contender, many manufacturers selected field buses instead. As introduced in previous work \cite{DBLP:journals/corr/abs-1804-07643}, we are starting to observe a change though. With the arrival of the `Time Sensitive Networking' (TSN) standards, Ethernet is expected to gain wider adoption for real-time robotic applications. There are currently several communication technologies based on the Ethernet protocols. Some protocols such as Profinet RT \cite{profinet} or Powerlink \cite{powerlink} use a network stack specifically designed for the protocol\footnote{These network stacks have been specifically designed to meet the desired real-time capabilities.}. Other protocols, such as the Data Distributed Services (DDS) \cite{dds}, OPC-UA \cite{opcua} or Profinet are built on top of the well known TCP/IP and UDP/IP OSI layers. This facilitates the process of interoperating with common transport protocols and ensures a high compatibility between devices. However, their corresponding network stacks and drivers are typically not optimized for real-time communications. The real-time performance is limited when compared to other Ethernet alternatives with specific network stacks.\\
% 
%This make these protocols dependent not only on the chosen real-time capabilities of operating system but also on the capabilities of its networking stack. Using  \\ \todo{Justify here our interest in the evaluation UDP of for higher level communications that use UDP as the base of the transport layer. For example ROS 2 with DDS.}

In this work, we aim to measure and evaluate the real-time performance of the Linux network subsystem with the UDP and IP protocols. UDP is used by several real-time transport protocols such as the Real Time Publish Subscribe protocol (RTPS). We aim to determine which configuration provides better isolation in a mixed-critical traffic scenario. To achieve real-time performance, the network stack will be deployed in a Real-Time Operating System (RTOS). In the case of Linux, the standard kernel does not provide real-time capabilities. However, with the Real-time Preemption patch (PREEMPT-RT), it is possible to achieve real-time computing capabilities as demonstrated \cite{JoC}. Despite the Linux network subsystem not being optimized for bounded maximum latencies, with this work, we expect to achieve reasonable deterministic communications with PREEMPT-RT, as well as a suitable configuration.\\

The content below is structured as follows: section \ref{related_work} introduces related previous work. In particular, we review the real-time performance of Linux with PREEMPT-RT and the real-time communications in Linux for robotics. Section \ref{background} outlines an overview of the key components of the Linux kernel we have used to configure our system to optimize the real-time communications over Ethernet. Section \ref{setup_results} discusses the experimental results obtained while evaluating the presented hypotheses. Finally, Section \ref{conclusions}  presents our conclusions and future work.\\

\section{Related work}
\label{related_work}

In \cite{Abeni_investigatingthe}, Abeni et al. compared the networking performance between a vanilla kernel 3.4.41 and the same version with PREEMPT-RT. They investigated the throughput and latency with different RT patch implementations. They also explained some of the key ideas of the Linux Network architecture in PREEMPT-RT and showed an evaluation of the network latency by measuring the round-trip latency with the UDP protocol. The results showed that the differences between the kernel version tested appeared when the system was loaded. For the kernel versions where the soft Interrupt Requests (IRQ) were served in the IRQ threads the results showed that latency was bounded. However, the influence of concurrent traffic was not discussed in the paper.\\ 
 
Khoronzhuk and Valris \cite{linaro_net_perf} showed an analysis of the deterministic performance of the network stack with the aim of using it with TSN. They used a real-time kernel 4.9 with a 1 Gbps Ethernet network interface. They concluded that there is jitter in the order of hundreds of microseconds across the networking stack layers, both in transmission and reception paths.\\ 

%We followed good real-time programming guides [4, 17, 19] and the experimental result will be presented
For an appropriate setup of our infrastructure, we made active use of the work presented at \cite{hard_soft_alison} and \cite{osadl_udp}, which provide some guidelines to configure a PREEMPT-RT based OS for real-time networking.
 
% Investigating the Network Performance of a Real-Time Linux
% Kernel : http://citeseerx.ist.psu.edu/viewdoc/download?doi=10.1.1.702.7571&rep=rep1&type=pdf

% Evaluation of Linux rt-preempt for embedded industrial devices for Automation and Power Technologies - A Case Study
% http://linuxdevices.io/ldfiles/article081/Sampath.pdf

% Realtime Performance of Networking Stack
% http://connect.linaro.org/resource/sfo17/sfo17-209/

% Othe protocols...

% Real-time CORBA performance on Linux-RT PREEMPT
% https://www.osadl.org/fileadmin/events/rtlws-2007/Traut.pdf

% Real-Time Performance Analysis in Linux-Based Robotic Systems
% https://pdfs.semanticscholar.org/8368/bede281fb89ad1b5908e240cf79150323759.pdf
% EtherCAT...

% It seems there is ongoing work such as XDP and AF_PACKET that promises to provide lower jitter and bounded latency. Some TSN features such as TAS support are being integrated in the mainline. We are interested in testing these new features in future work.

\section{Setting up real-time communications in Linux}
\label{background}

%\subsection{The Linux Networking Architecture}
%\label{sub}

% Main components of the The Linux Networking stack.

% Transmission data path

% Reception data path
% - NAPI
% - Network IRQ threads.
% - Multiqueue devices.
% - Ingress Policing: filter

%In this section we are going to explain some of the key parts of the networking stack and which are the configurations we have used for the experiments. In particular, we will try to find the best possible configuration for deterministic Ethernet communications for a multi-queue network interface with mixed critical traffic. \\ 

%The system must be configured correctly for deterministic communications, just by running an application under Preempt-RT does not provide deterministic behavior to the application. This imply having a good understanding of the real-time application and the kernel internals, this configuration can be sometimes complex for the user. Some of these configurations are for example setting the thread priorities, bind some threads and IRQs to some CPUs if is required, lock the memory, etc. In this work we are focusing on applications that require real-time communications so we need to have a general understanding about the Linux Networking architecture and how it works under Preempt-RT.\\

\subsection{The Real-time Preemption patch (PREEMPT-RT)}
\label{preempt-rt}

There are currently different approaches to use Linux for real-time applications. A common path is to leave the most critical tasks to an embedded RTOS and give to Linux the highest level commands. A second approach is to use a dual-kernel scheme like Xenomai \cite{xenomai} and RTAI \cite{rtai} which deploy a microkernel running in parallel with a separate Linux kernel. The problem for this kind of solution is that it requires special tools and libraries.\\

A third approach is to use a single-kernel. The Real-Time Linux (RTL) Collaborative Project \cite{rtl} is the most relevant open-source solution for this option. The project is based on the PREEMPT-RT patch and aims to create a predictable and deterministic environment turning the Linux kernel into a viable real-time platform. The ultimate goal of the RTL project is to mainline the PREEMPT-RT patch. The importance behind this effort is not related to the creation of a Linux-based RTOS, but to provide the Linux kernel with real-time capabilities. The main benefit is that it is possible to use the Linux standard tools and libraries without the need of specific real-time APIs. Also, Linux is widely used and strongly supported, this helps to keep the OS updated with new technologies and features, something which is often a problem in smaller projects due to resource limitations. \\

%\todo{mention POSIX}

%It is widely used in robotics \cite{abb}, telecom, manufacturing, and medical industries.\\

%The standard Linux kernel only meets soft real-time requirements but there are several real-time exten- sions available for Linux right now. One of them is the Realtime Preemtpion Patch (RT-Preempt) devel- oped by Ingo Molnar. Unlike other Linux real-time extensions RT-Preempt doesn’t use a micro-kernel but brings hard real-time capabilities directly into the Linux kernel. The big advantage of this solu- tion is that the user can use his standard linux tools for development, using the POSIX API for his ap- plications and doesn’t need to lern special real-time APIs.

\subsection{The Linux Networking Architecture}
\label{LNS}

While it is possible to bypass the Linux Network Stack using custom drivers or user-space network libraries, we are interested in using the Linux Network Stack; mainly, because it is easier to maintain and integrate with a user-space application or communication middlewares. In addition, the Linux Network Stack supports a wide range of drivers which allow to deploy an application in different devices.\\

%The Linux Network Stack is a complex subsystem, it is not straightforward to understand how to configure the system to improve determinism. The transmission and reception processes work in different ways and the execution context may vary depending on the network load. 

%traffic policing and traffic shaping

\subsubsection{Linux Traffic Control}
\label{tc}

An important module of the networking subsystem is the Linux kernel packet scheduler, which is configured with the user-space tool Linux Traffic Control (TC)  \cite{tc}. TC provides mechanisms to control the way enqueued packets are sent and received, it provides a set of functionality such as shaping, scheduling, policing and dropping network traffic. \\

% It offers a large set of functionality for classifying, scheduling, policing, and dropping network traffic

The main element of the Linux packet scheduler module are the queuing disciplines (Qdisc), which are network traffic disciplines to create queues and quality of service (QoS) rules for reception and transmission. There are ingress and egress Qdisc for reception and transmission respectively. The egress Qdisc provides shaping, scheduling and filter capabilities for data transmission from the network protocol layers to the network interface ring buffers. On the other hand, the ingress Qdisc provides filter and dropping capabilities for the reception path from the network interface ring buffers to the network protocol layers (although these are commonly less used).\\

%Queuing disciplines are commonly used as attempts to compensate for various networking conditions, like reducing the latency for certain classes of network packets, and are generally used as part of quality of service (QoS) measures

% This tool can be used to prioritize traffic and lower the latency of critical traffic.
% Which Qdisc are suitable for rt?

For the egress Qdisc there are two basic types of disciplines: classless Qdisc and classful Qdisc. The classless Qdisc does not contain another Qdisc so there is only one level of queuing. The classless Qdisc only determines whether the packet is classified, delayed or dropped. The classful Qdisc can contain another Qdisc, so there could be several levels of queues. In such case, there may be different filters to determine from which Qdisc packets will be transmitted.\\

\begin{figure}[h!]
\centering
 \includegraphics[width=0.5\textwidth]{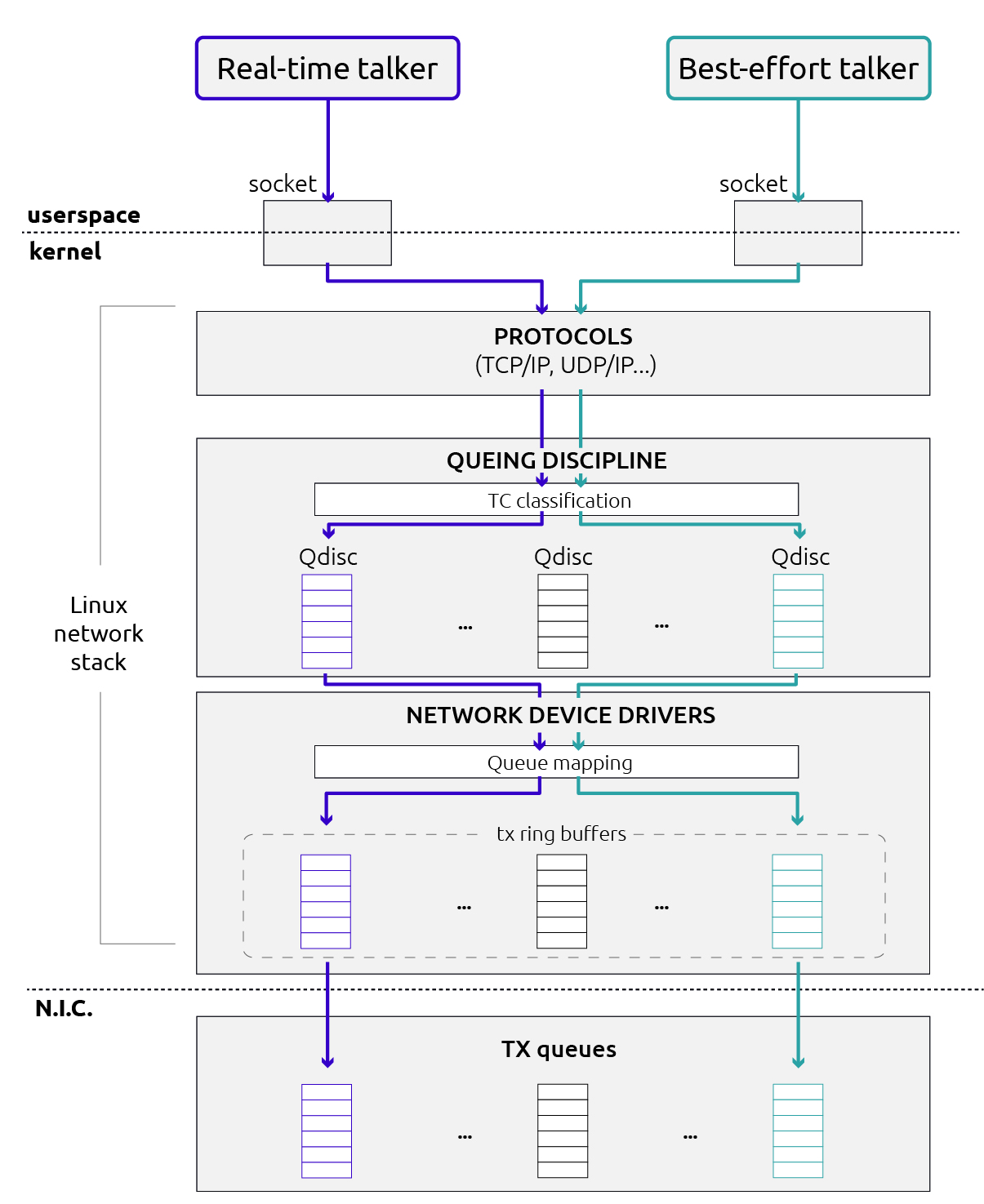}
\caption{\footnotesize Linux networking stack transmission path}
\label{tx_stack}
\end{figure}

Qdisc can be used to avoid traffic congestion with non real-time traffic at the transmission path (figure \ref{tx_stack}). For a classless Qdisc, the default discipline is the PFIFO\_FAST, which has three FIFO priority bands. In the case of a classful Qdisc there is the PRIO qdisc which can contain an arbitrary number of classes of differing priority. There are also specific egress Qdisc destined for multiqueue network devices, for example the MQPRIO Qdisc \cite{mqprio}, which is a queuing discipline to map traffic flows to hardware queues based on the priority of the packet. This Qdisc will dequeue packets with higher priority allowing to avoid contention problems in the transmission path. In addition to the priority Qdisc, it is common to attach a shaper to limit low priority traffic bandwidth such as a the `Token Bucket Filter' TBF Qdisc \cite{tbf}.\\

% Mention: HTB (Hierarchical Token Bucket)

%If a basic FIFO queue is used for the egress Qdisc low priority traffic can block higher priority traffic increasing the latency in the transmission path. To prevent this it is possible to create an egress Qdisc allows to create queue packets in different queues according to their priority.

% by attaching a shaper (for example, tc-tbf(8) to these bands to make sure they cannot dominate the link.
% Token Bucket Filter

Recently, because of the interest in support TSN in the Linux network stack, new Qdiscs have been created or are currently under development. The IEEE 802.1Q-2014 Credit Based Shaper (CBS) \cite{cbs}, Qdisc has already been included from kernel 4.15. The CBS is used to enforce a Quality of Service by limiting the data rate of a traffic class. Currently there are two Qdisc under development, the `Earliest Transmit Time First (ETF)' \cite{lwn_etf} which provides a per-queue transmit time based scheduling and the `Time-Aware Priority Scheduler' (TAPRIO) which provides per-port scheduling. These Qdisc will allow to create deterministic scheduling in software or to offload the work to the network hardware if it is supported.\\

%\todo{Explain some Qdisc being developed for TSN: SO\_TXTIME, etf,  taprio and the cbs qdisc already included.}

%This work consists of a set of kernel interfaces that can be used by applications that require (time-based) Scheduled Tx of packets.
% It is comprised by 3 new components to the kernel:

% - SO_TXTIME: socket option + cmsg programming interfaces.

%  - etf: the "earliest txtime first" qdisc, that provides per-queue
%	 TxTime-based scheduling. This has been renamed from 'tbs' to
%	 'etf' to better describe its functionality.

%  - taprio: the "time-aware priority scheduler" qdisc, that provides
%	    per-port Time-Aware scheduling;

\subsubsection{Traffic classification}
\label{cl}
In order to steer a traffic flow to a Qdisc or to a ring buffer, the traffic flow must be classified usually by marking the traffic with a priority. There are several ways to set the priority of a specific traffic flow: a) from the user-space using socket options SO\_PRIORITY and IP\_TOS, b) with iptables and c) with net\_prio cgroups. Setting the priority of a flow maps the traffic from a socket (or an application) to a Socket Buffer (SKB) priority, which is an internal priority for the kernel networking layer. The SKB priority is used by the MQPRIO Qdisc to map the traffic flow to a traffic class of the Qdisc. At the same time, each traffic class is mapped to a TX ring buffer. \\

% Other possible configurations.
% Limit max rate per queue.
% Shaping
% Filter/dropping

\subsubsection{Network hard IRQ threads and softirqs}
\label{irqs}

At the reception path, the processing of the packets is driven by the kernel interrupt handling mechanism and the ``New API'' (NAPI) network drivers. NAPI is a mechanism designed to improve the performance for high network loads. When there is a considerable incoming traffic, a high number of interrupts will be generated. Handling each interrupt to process the packets is not very efficient when there are many packets already queued. For this reason NAPI uses interrupt mitigation when high bandwidth incoming packets are detected. Then, the kernel switches to a polling-based processing, checking periodically if there are queued packets. When there is not as much load, the interrupts are re-enabled again. In summary, the Linux kernel uses the interrupt-driven mode by default and only switches to polling mode when the flow of incoming packets exceeds a certain threshold, known as the ``weight'' of the network interface. This approach works very well as a compromise between latency and throughput, adapting its behavior to the network load status. The problem is that NAPI may introduce additional latency, for example when there is bursty traffic.\\

%The networking subsystem uses two software interrupts (softirq), one each for transmit and receive processing. The receive softirq is where NAPI processing (the polling of interfaces for new packets) is done. If there are a lot of packets to handle, this processing can take a long time. 

%As seen NAPI moves processing to softirq level. Linux uses the ksoftirqd as the general solution to schedule softirq's to run before next interrupt and by putting them under scheduler control. Also this prevents consecutive softirq's from monopolizing the CPU. 

%The softirq calls your poll function to process incoming packets until you have no more packets or until the NAPI "budget" is exhausted. (In the latter case, the softirq is later re-invoked by the ksoftirqd thread [I think].)

%Essentially, a software interrupt is meant to look like a hardware interrupt, except that it runs at a lower priority.

There are some differences between how PREEMPT-RT and a normal kernel handle interrupts and consequently how packets are handled at the reception path. The modifications of PREEMPT-RT allow to configure the system to improve the networking stack determinism.\\ 

In PREEMPT-RT, most interrupt request (IRQ) handlers are forced to run in threads specifically created for that interrupt. These threads are called IRQ threads \cite{lwn_irq_threads}. Handling IRQs as kernel threads allows priority and CPU affinity to be managed individually. IRQ handlers running in threads can themselves be interrupted, so that the latency due to interrupts are mitigated. For a multiqueue NIC, there is an IRQ for each TX and RX queue of the network interface, allowing to prioritize the processing of each queue individually. For example, it is possible to use a queue for real-time traffic and raise the priority of that queue above the other queue IRQ threads.\\

Another important difference is the context where the softirq are executed. From version 3.6.1-rt1, the soft IRQ handlers are executed in the context of the thread that raised that Soft IRQ \cite{lwn_softirq}. This means that the NET\_RX soft IRQ will be normally executed in the context of the network device IRQ thread, which allows a fine control of the networking processing context. However, if the network IRQ thread is preempted or it exhausts its NAPI weight time slice, it is executed in the ksoftirqd/n (where n is the logical number of the CPU). \\ 

Processing packets in the ksoftirqd/n context is troublesome for real-time because this thread is used by different processes for deferred work and can add latency. Also, as the ksoftirqd runs with SCHED\_OTHER policy, the thread execution can be easily preempted. In practice, the soft IRQs are normally executed in the context of NIC IRQ threads and in the ksoftirqd/n thread for high network loads and under heavy stress (CPU, memory, I/O, etc..).\\

\subsubsection{Socket allocation}
\label{skb_alloc}

One of the current limitations of the network stack for bounded latency is socket memory allocation. Every packet in the network stack needs a \emph{sckbuff} struct which holds meta-data of the packet. This struct needs to be allocated for each packet, and the time required for the allocation represents a large part of the overhead for processing the packet and jitter source.\\

One of the last projects of the Linux network developers is the XDP or eXpress Data Path \cite{xdp} which aims to provide a high performance, programmable network data path in the Linux kernel. XDP will provide faster packet processing by eliminating the socket meta-data allocation. Despite real-time communications are not the main motivation behind this project, XDP looks like an interesting feature to be used as an express data path for real-time communications \cite{jesus_tsn}.\\

\section{Experimental setup and results}
\label{setup_results}

%In our evaluation we analyzed the lowest POWER- LINK cycle times which could be achieved on a Linux POWERLINK MN with the current openPOWER- LINK stack and how much system load it generates. Additionally we measured the quality of the POW- ERLINK timing on the network.

To evaluate the real-time performance of the network stack, we used two embedded devices, measuring the latencies of a round-trip test.\\ 

%To get the worst case values, all the tests in the paper have been done with heavy system load.

% Robotic use case: 

\subsection{Round-trip test}
\label{setup}

The network latency is measured as the round-trip time (RTT), also called ping-pong test. For the test we use a client in one of the devices and a server in the other. The round-trip latency is measured as the time it takes for a message to travel from the client to the server, and from the server back to the client. 

For the client and server one we used a modified version of cyclictest \cite{cyclictest} which allows us to save the statistics and create latency histograms which show the amount of jitter and worst-case latency of the test. Additionally, we count the number of missed deadlines for a 1 millisecond target loop time.\\

For the timing loop we used the \emph{clock\_nanosleep} primitive. We also used memory locking, the FIFO scheduler and set a real-time priority of 80. In all the tests, we marked the traffic as priority traffic using the socket option SO\_PRIORITY. To generate load in the system, we used the program \emph{stress} and, to generate traffic, the program \emph{iperf}.\\
%In the last case, rt-isolation, we use \emph{cgroups} to use the CPU 1 for the real-time programs. We bind also all the IRQs to CPU 0 except the IRQs of the priority queue. \\

\begin{figure}[h!]
\centering
 \includegraphics[width=0.4\textwidth]{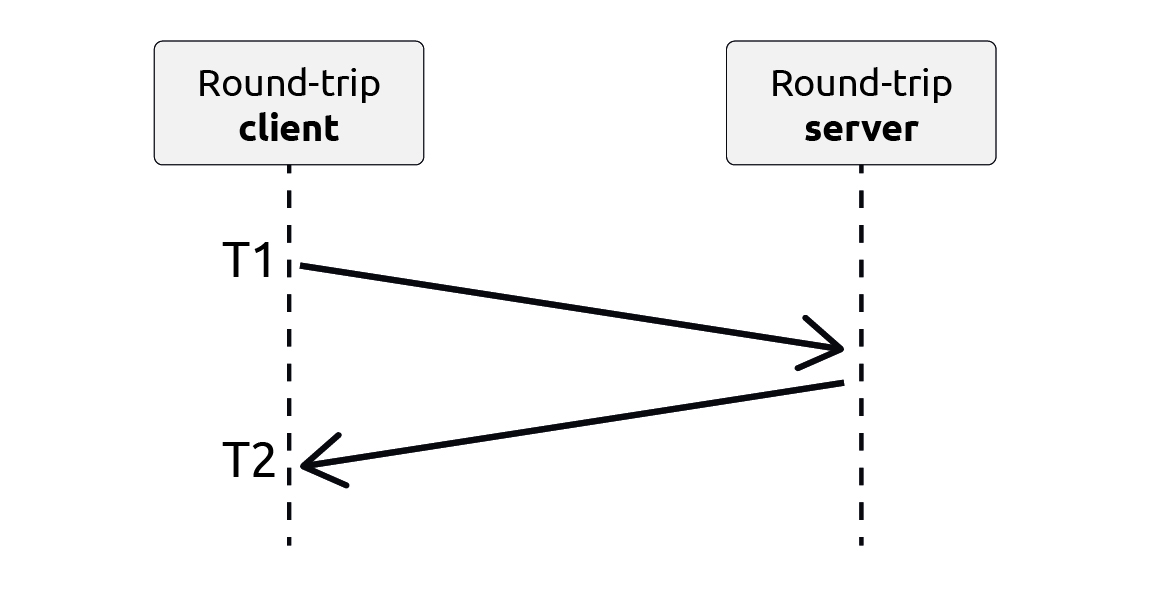}
\caption{\footnotesize Graphical presentation of the measured round-trip latency. T1 is the time-stamp when data is send from the round-trip client and T2 is the time-stamp when data is received again at the round-trip client. Round-trip latency is defined as T2 - T1.}
\label{ping-pong}
\end{figure}

\subsubsection{Experimental setup}

%For the setup we use two embedded devices with ZYNQ-7020 chipsets,

The main characteristics of the embedded device are:
\begin{itemize}
    \item Processor: ARMv7 Processor (2 cores).
    %\item \todo{OS: OpenEmbedded (poky).}
    \item Kernel version: 4.9.30.
    \item PREEMPT-RT patch (when used): rt21. 
    \item Link capacity: 100/1000 Mbps, Full-Duplex.
    % Cyclictest maximum latency?
\end{itemize}

%which have a Dual-core ARM Cortex-A9 MPCore processor with a 667 MHz base frequency. We use the following kernel versions 4.9.30 and 4.9.30-rt21, in the following referred as no-RT and RT respectively.

Each device has a switched endpoint as network interface. The devices are connected in linear topology with a PC. The PC is simply used to generate contending traffic, also referred in this work as background traffic. \\

%\begin{figure}[h!]
%\centering
% \includegraphics[width=0.4\textwidth]{images/Scenario1_4.jpg}
%\caption{\footnotesize Experimental setup networking devices overview.}
%\label{scenario}
%\end{figure}

The traffic used to measure the end-to-end latency is marked with higher priority than the background traffic. The embedded device network interface has 3 queues (tx\_q0, tx\_q1, tx\_q3). In our case, the network drivers use the Priority Code Point (PCP) of the Ethernet header to map a frame with a TX queue. Priority 4 is mapped to tx\_q0, priorities 2 and 3 to tx\_q1 and the rest of priorities to tx\_q2. We configure a MQPRIO qdisc to map the SKB priorities with the same classification criteria than the one used by the network device drivers. The Qdisc queues are mapped one to one with each hardware queue, so we can avoid traffic contention at the network layer in the transmission path.\\

% the real-time traffic can be isolated from other traffic.

\begin{lstlisting}
tc qdisc replace dev eth1 root  mqprio num_tc 3 \
map 2 2 1 1 0 2 2 2 2 2 2 2 2 2 2 2  
queues 1@0 1@1 1@2 hw 0
\end{lstlisting}

This command maps the SKB priority 4 with queue 0, SKB priorities 2 and 3 with queue 1 and the rest of priorities with queue 2. Checking the classes setting, we get the following output:

\begin{lstlisting}
root@DUT1:~# tc -g qdisc show dev eth1
qdisc mqprio 8001: root  tc 3 map 2 2 1 1 0 2 2 2 2 2 2 2 2 2 2 2
             queues:(0:0) (1:1) (2:2)
qdisc pfifo_fast 0: parent 8001:3 bands 3 priomap  1 2 2 2 1 2 0 0 1 1 1 1 1 1 1 1
qdisc pfifo_fast 0: parent 8001:2 bands 3 priomap  1 2 2 2 1 2 0 0 1 1 1 1 1 1 1 1
qdisc pfifo_fast 0: parent 8001:1 bands 3 priomap  1 2 2 2 1 2 0 0 1 1 1 1 1 1 1 1
\end{lstlisting}

Apart from the networking layer mapping, we need to map the SKB priorities with the priority used by the Ethernet layer, which is the PCP. We set a one to one configuration. This sets the priority in VLAN field of the Ethernet header according to the SKB priority configured in the user-space. The PCP is used by the Network interface Card (NIC) to map traffic to the hardware queues. 

\begin{lstlisting}
root@DUT1:~# ip link set eth1.2 type vlan egress 0:0 1:1 2:2 3:3 4:4 5:5 6:6 7:7
\end{lstlisting}

%TC also allows to configure Qdisc for ingress traffic to create ingress policing rules. We are not going to use it for the moment.

\subsection{Task and IRQ affinity and CPU shielding}
\label{affinity}

%\cite{cpu_part}

% CPUs can be partitioned to separate the resources of tasks and interrupts with different focus. In a real time system, CPU partitioning can be used to separate CPUs dedicated to real time tasks and their corresponding interrupts.

In real-time systems, real-time tasks and interrupts can be pinned to a specific CPU to separate their resources from non real-time tasks. This is an effective way to prevent non real-time processes interferences. \\

There are several ways to set the affinity of tasks and IRQs with CPUs. For the experiments, we decided to compare two levels of isolation. In the first case, we pin the real-time task the IRQ of the real-time traffic queue to the same CPU. We use ``pthread\_setaffinity\_np'' and ``smp irq affinity'' to set the priorities of the IRQs.\\

% cpusets cgroups. a mechanism to assign a set of CPUs and NUMA memory nodes (if NUMA is available) to a set of tasks.
In the second case, we use \emph{cpusets} \cite{cpuset}, which is part of the Linux \emph{cgroups} to assign a CPU for real-time tasks. With this method, we can also migrate all processes running in the isolated CPU, so that only the real-time task is allowed to run in that CPU\footnote{Only some kernel processes will run in the isolated CPU which will reduce the system jitter considerably.}. We also set the affinity of all the IRQs to the non real-time CPU, while the IRQs of the real-time queue (of the network device) are set with affinity in the isolated CPU.\\

%The second method uses isolcpus to isolate a CPU for real-time applications. This method allows to achieve a better isolation but it imply a more complex configuration. Using isolcpus we prevent the scheduler to schedule any task in the isolated CPU. We also set the affinity of all the IRQs to the non real-time CPU and we set only the affinity to the isolated CPU for the IRQs of the real-time queue of the network device.\\

In the experimental setup, we use the described methods to isolate applications sending and receiving real-time traffic. 

The tests are run using different configurations: \texttt{no-rt}, \texttt{rt-normal}, \texttt{rt-affinities} and \texttt{rt-isolation}. In the first case, \texttt{no-rt}, we use a vanilla kernel. In the second case, \texttt{rt-normal}, we use a PREEMPT-RT kernel without binding the round-trip programs and network IRQs to any CPU. In the third case, \texttt{rt-affinities}, we bind the IRQ thread of the priority queue and the client and server programs to CPU 1 of each device. Finally, in the fourth case, \texttt{rt-isolation}, we run the round-trip application in an isolated CPU. In all the cases, we set the priority of the RTT test client and server to a 80 value.\\

In order to have an intuition about the determinism of each configuration, we ran a 1 hour cyclictest, obtaining the following worst-case latencies: no-rt: 13197 $\mu s$, rt-normal/rt-affinities: 110 $\mu s$ and rt-isolation: 88 $\mu s$. \\

\subsubsection{System and network load}

For each case, we run the tests in different load conditions: idle, stress, tx-traffic and rx-traffic:

\begin{itemize}
    \item \texttt{idle}: No other user-space program running except the client and server.
    \item \texttt{stress}: We generate some load to stress the CPU and memory and block memory.\footnote{Command used: stress -c 2 -i 2 -m 2 --vm-bytes 128M -d 2 --hdd-bytes 15M}
    \item \texttt{tx-traffic}: We generate some concurrent traffic in the transmission path of the client. We send 100 Mbps traffic from the client to the PC. 
    \item \texttt{rx-traffic}: We generate some concurrent traffic in the reception path of the server. We send 100 Mbps traffic from the PC to the server.
\end{itemize}

When generating concurrent traffic, there is also congestion in the MAC queues of both devices. However, as the traffic of the test is prioritized, the delay added by the link layer is not meaningful for the test.\\ 

%For all the measurements we run the test during X hrs. While it is possible to observe the effect of the system load and concurrent traffic in short test some latency appear randomly when a series of %events happens at the same time. For this reason all the real-time tests should be long term measurements.\\

%\subsection{Measuring the TX and RX path latency}
%\label{onewaytest}

%The round-trip test give as an intuition about the delays along the network stack path in total. However, we cannot measure the transmission and reception delays separately. To break down the latency in the transmission and reception paths we need to timestamp the packet at when entering or going out of the network stack. As the hardware used for the experimental setup has hardware timestamping capabilities we are going to use these timestamps to measure the delays at the transmission and reception. To have coherent timestamping sources we need to synchronize the system clock (CLOCK\_REALTIME) with the PTP clock of the network device first. For the synchronization we use ptp4l and phc2sys.\\

%We repeat the same measurements than in the round-trip case. Now we can observe in more detail the network stack latency.\\

\subsection{Results}
\label{results}

We have compared the results obtained for a round-trip test of 3 hours of duration, sending UDP packets of 500 Bytes at a 1 millisecond rate. The Tables \ref{nort-table}, \ref{rt-table}, \ref{affinity-table}, \ref{isolated-table} show the statistics of the different configuration used under different conditions. For real-time benchmarks, the most important metrics are the worst-cases (Max), packet loss and the number of missed deadlines. In this case, we decided to set a 1 millisecond deadline to match it with the sending rate.\\ %\todo{ and the Packets loss}.\\

As it can be seen in Table \ref{nort-table}, the non real-time kernel results in what seems to be the best average performance, but in contrast, it has a high number of missed deadlines and a high maximum latency value; even when the system is idle. The latency suffers specially when the system is stressed due to the lack of preemption in the kernel. \\

For \texttt{rt-normal} (Table \ref{rt-table}), the latencies are bounded when the system is stressed. When generating concurrent traffic, we observe higher latency values and some missed deadliness.\\ 

For \texttt{rt-affinities}, we can see an improvement compared to the previous scenario. Specially for concurrent traffic (Table \ref{affinity-table}). We can also see that when pinning the round-trip threads and the Ethernet IRQs for the priority to the same CPU, the latency seems to be bounded.\\

\begin{figure}[h!]
\centering
 \includegraphics[width=0.45\textwidth]{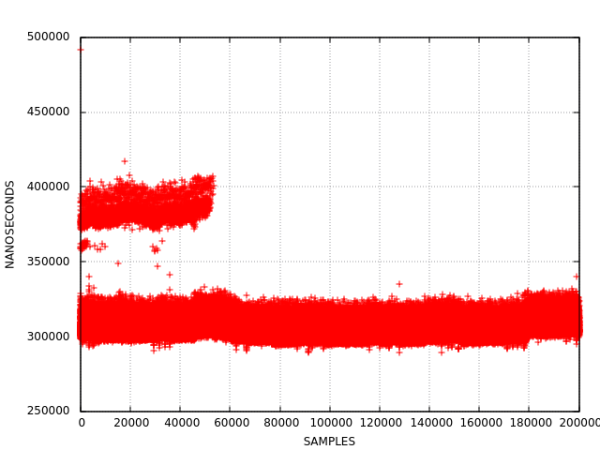}
\caption{\footnotesize Time-plot for isolated CPU. At the beginning, we can observe the effect of the scheduler ticker preempting the real-time task and adding latency to the round-trip test latencies.}
\label{plot_iso_ticker}
\end{figure}

In the case of \texttt{no-isolation} (table \ref{isolated-table}), we appreciate a similar behavior when compared to the affinity case. We can see that stressing the non isolated CPU has some impact on the tasks of the isolated core. However, in the idle case, for short test we observed very low jitter. To the best of our knowledge, one of the main contributions of such latency was the scheduler ticker, which is generated each 10 milliseconds\footnote{This effect was observed using kernel tracer `ftrace' and correlating the latency measurements with the traces during the test.}. While it is possible to avoid it in the client because it runs in a timed loop, in the server side, it is not possible to avoid the ticker. As both devices are not synchronized, at some point, the clock of the server side drifts from the client and the scheduler ticker interferes in the server execution. This effect can be seen in Figure \ref{plot_iso_ticker}.\\ 

When running the test with 200 Mbps RX-traffic, we observed that best-effort traffic is processed in the ksoftirqd/0 context continuously. This generates high latency spikes in all the cases, even for the isolation case. To trace the source of these latency spikes, we should trace the kernel taking a snapshot when the latency occurs. \\

%                   & Min($\mu$s)   & Avg($\mu$s) & Max($\mu$s)  & \multicolumn{1}{c|}{missed deadlines samples} \\ \hline
\begin{table*}[ht]
\centering
\caption{Round-trip latency results: No RT}
\label{nort-table}
\begin{tabular}{|c|c|c|c|r|c|}
\hline
\multicolumn{6}{|c|}{No RT, Kernel version: 4.9.30}                       \\ \hline
                  & Min($\mu$s)    & Avg($\mu$s) & Max($\mu$s) & \multicolumn{1}{c|}{Missed deadline}   & Packet loss \\ \hline
Idle              & 193  & 217  & 1446 & 15 / 600000    & 0 / 600000          \\ \hline
Stress            & 262  & 560 & 46742 & 20979 / 600000 & 0 / 600000      \\ \hline
TX traffic at 100 Mbps & 195   & 378 & 7160 & 2298 / 600000  & 0 / 600000                      \\ \hline
RX traffic at 100 Mbps & 192   & 217 & 1426 & 22 / 600000    & 0 / 600000                        \\ \hline
\end{tabular}
\end{table*}

\begin{table*}[ht]
\centering
\caption{Round-trip latency results: RT Normal}
\label{rt-table}
\begin{tabular}{|c|c|c|c|r|c|}
\hline
\multicolumn{6}{|c|}{RT Normal, Kernel version: 4.9.30-rt21}                     \\ \hline
                  & Min($\mu$s)   & Avg($\mu$s) & Max($\mu$s)  & \multicolumn{1}{c|}{Missed deadline}   & Packet loss \\ \hline
Idle              & 251    & 266 & 522 & 0 / 600000 & 0 / 600000                      \\ \hline
Stress            & 254 & 341  & 618  & 0 / 600000  & 0 / 600000                        \\ \hline
TX traffic at 100 Mbps & 265 & 320 & 25727  & 20 / 600000   & 0 / 600000                       \\ \hline
RX traffic at 100 Mbps & 263 & 292  & 898   & 9 / 600000    & 0 / 600000                        \\ \hline
\end{tabular}
\end{table*}

\begin{table*}[ht]
\centering
\caption{Round-trip latency results: RT, CPU Affinity}
\label{affinity-table}
\begin{tabular}{|c|c|c|c|r|c|}
\hline
\multicolumn{6}{|c|}{RT Affinity, Kernel version: 4.9.30-rt21}          \\ \hline
                  & Min($\mu$s) & Avg($\mu$s) & Max($\mu$s) & \multicolumn{1}{c|}{Missed deadline}   & Packet loss \\ \hline
Idle              & 275 & 289  & 644 & 0 / 600000   & 0 / 600000                       \\ \hline
Stress            & 277 & 358 & 828  & 0 / 600000   & 0 / 600000                        \\ \hline
TX traffic at 100 Mbps & 277 & 322  & 568 & 0 / 600000  & 0 / 600000                        \\ \hline
RX traffic at 100 Mbps & 274 & 287  & 592  & 0 / 600000 & 0 / 600000                        \\ \hline
\end{tabular}
\end{table*}

\begin{table*}[ht]
\centering
\caption{Round-trip latency results: RT, CPU Isolated}
\label{isolated-table}
\begin{tabular}{|c|c|c|c|r|c|}
\hline
\multicolumn{6}{|c|}{RT, CPU Isolated, Kernel version: 4.9.30-rt21}     \\ \hline
                  & Min($\mu$s) & Avg($\mu$s) & Max($\mu$s) & \multicolumn{1}{c|}{Missed deadline}   & Packet loss \\ \hline
Idle              & 297 & 311 & 592  & 0 / 600000   & 0 / 600000                         \\ \hline
Stress            & 298 & 355 & 766 & 0 / 600000    & 0 / 600000                         \\ \hline
TX traffic at 100 Mbps & 301 & 336 & 617 & 0 / 600000   & 0 / 600000                        \\ \hline
RX traffic at 100 Mbps & 296 & 312 & 542 & 0 / 600000   & 0 / 600000                         \\ \hline
\end{tabular}
\end{table*}

\begin{figure*}[h!]
  \begin{subfigure}[t]{.5\textwidth}
    \centering
    \includegraphics[width=0.8\linewidth]{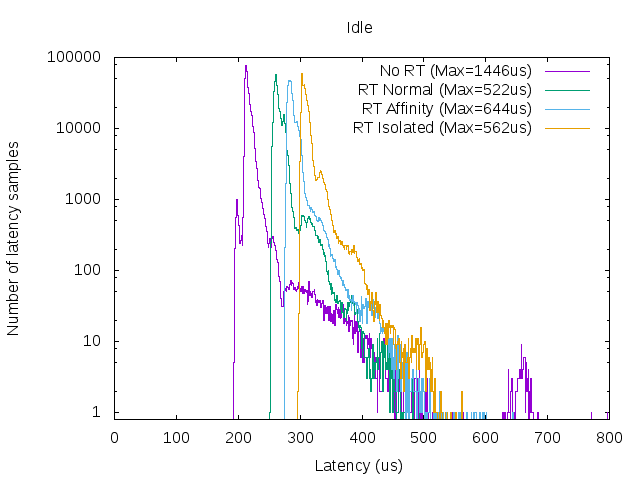}
    \caption{ }
    \label{plot_idle}
  \end{subfigure}
  \hfill
  \begin{subfigure}[t]{.5\textwidth}
    \centering
    \includegraphics[width=0.8\linewidth]{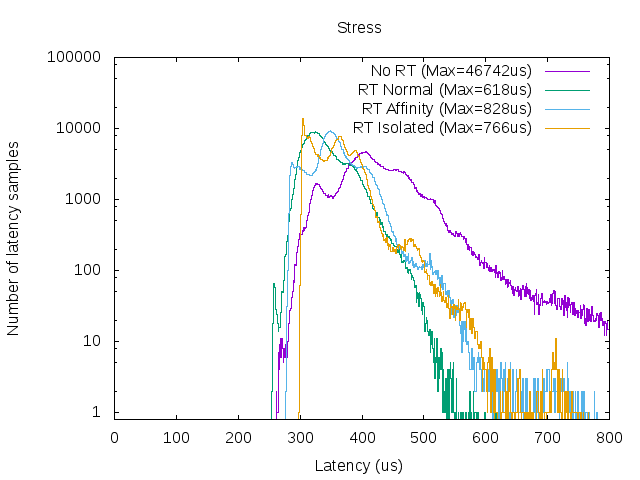}
    \caption{}
    \label{plot_stress}
  \end{subfigure}

  \medskip

  \begin{subfigure}[t]{.5\textwidth}
    \centering
    \includegraphics[width=0.8\linewidth]{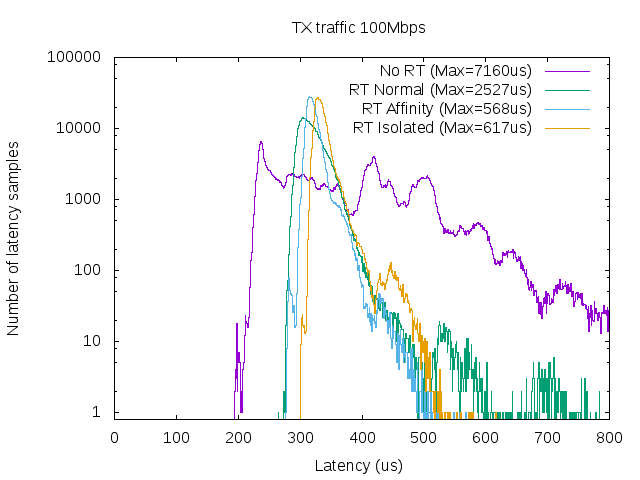}
    \caption{}
    \label{plot_tx}
  \end{subfigure}
  \hfill
  \begin{subfigure}[t]{.5\textwidth}
    \centering
    \includegraphics[width=0.8\linewidth]{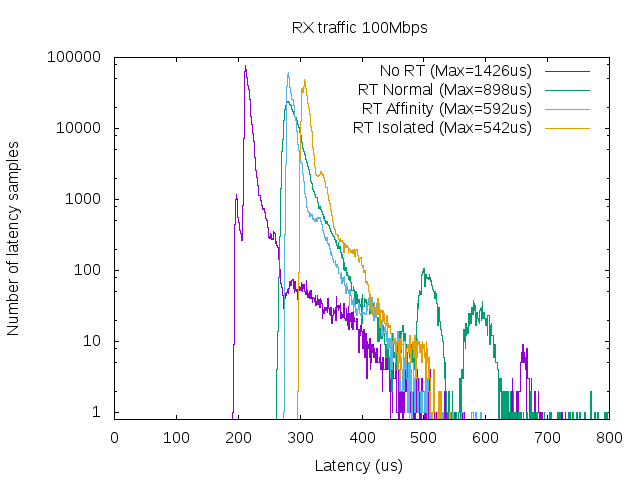}
    \caption{}
    \label{plot_rx}
  \end{subfigure}
  \caption{ \footnotesize Real-time Ethernet round-trip-time histograms. a) Idle system.  b) System under load (stress).  c) Concurrent low priority traffic in the transmission path.  d) Concurrent low priority traffic in the reception path. }
\end{figure*}

%and under different load scenarios.\\

% Explain the setup
% - Two embedded boards with ZYNQ chipsets.
% - Version of the kernel
% - Kernel configuration (most relevant ones).
% - Connected point to point and a PC in daisy-chain.
% 
%
% Explain the benchmark program and measurements
% - Round-trip + cyclictest
% - The network interface prioritize marked packets so the delay is not meaningful for this test.
% - Stress
% - Iperf
%
% Introduce different configurations:
% - Normal
% - Affinities
% - Isolation

% Mention cyclictest result with and without stress. 

% Present different scenarios:
% - Idle
% - System load (Stress)
% - Tx Concurrent traffic: 100 Mbps, 200Mbps?
% - Rx Concurrent traffic: 100 Mbps, 200Mbps

% tick timers
% ptp

% Show results
% - table: avg, min, max, packet lost
% - plots 10min (substitute them later for 10hrs plots) 

% A common way to benchmark networks is to set up two computers and have a sender transmit a message to a receiver that echoes it back. That way the sender can measure the round-trip time (RTT) and gather statistics of the network. This generally works well, but large operating system stacks and device drivers can potentially
% The round-trip time is the difference between the receive and transmit timestamps. We also recorded the sequence number of each packet and the IP address of the receiver node in order to detect packet loss and track ordering.

%  measuring performance under heavy load is important to observe worst-case performance.
%
\section{Conclusion and future work}
\label{conclusions}

The results obtained prove that the Linux real-time setup presented improves greatly the determinism of communications using the UDP protocol. First, we confirm that the communication delay caused when the system is under heavy load is mitigated by making use of a real-time kernel and by running the application with real-time priority.

Second, we demonstrate that, whenever there is concurrent traffic, simply setting the priorities of the real-time process is not enough.  Separating the real-time application and the corresponding interrupt in a CPU seems to be an effective approach to avoid high latencies. For higher concurrent traffic loads, however, we can still see unbounded latency and further research is required to overcome this limitation with our current setup. \\
%we would like to do more research to find the sources of this and try to avoid it if possible.\\

We conclude that, under certain circumstances and for a variety of stress and traffic overload situations, Linux can indeed meet some real-time constraints for communications. Hereby, we present an evaluation of the Linux communication stack meant for real-time robotic applications. Future work should take into account that the network stack has not been fully optimized for low and bounded latency; there is certainly room for improvement. It seems to us that there is some ongoing work inside the Linux network stack, such as the XDP \cite{xdp} project, showing promise for an improved real-time performance. In future work, it might be interesting to test some of these features and compare the results.\\

\newpage
\bibliographystyle{IEEEtran}
\bibliography{references}

%\appendices
%\newpage

\onecolumn

%\section*{Appendix}
%\appendix

%\begin{thebibliography}{1}
%\end{thebibliography}

%\begin{biography}[{\includegraphics[width=1in,height=1.25in,clip,keepaspectratio]{mshell}}]{Michael Shell}

%\begin{IEEEbiography}[{\includegraphics[width=1in,height=1.25in,clip,keepaspectratio]{picture}}]{John Doe}
%\blindtext
%\end{IEEEbiography}


%\vfill


\end{document}